\documentclass[acmtocl]{acmtrans2m}

\title{Defining Functions on Equivalence Classes}
\author{LAWRENCE C. PAULSON\\
         University of Cambridge}

\begin{abstract}
A \emph{quotient construction} defines an abstract type from a concrete
type, using an equivalence relation to identify elements of the concrete
type that are to be regarded as indistinguishable. The elements of a
quotient type are \emph{equivalence classes}: sets of equivalent concrete
values. Simple techniques are presented for defining and reasoning about
quotient constructions, based on a general lemma library concerning
functions that operate on equivalence classes. The techniques are applied
to a definition of the integers from the natural numbers, and then to the
definition of a recursive datatype satisfying equational constraints.
\end{abstract}

\let\ts=\thinspace
\newcommand\lbrakk{\mathopen{[\![}}
\newcommand\rbrakk{\mathclose{]\!]}}

\usepackage{isabelle,isabellesym}
\usepackage{amsmath,amssymb}
\usepackage{url}

\category{G2.0}{Discrete Mathematics}{General}
            
            

\begin{document}
\begin{bottomstuff} 
\emph{Author's address}: Computer Laboratory, 15 JJ Thomson Avenue,
Cambridge CB3 0FD, England. \emph{E-mail}: \texttt{lcp@cl.cam.ac.uk}
\end{bottomstuff}

\maketitle

\section{Introduction}
 
Equivalence classes and quotient constructions are familiar to every
student of discrete mathematics. They are frequently used for defining
abstract objects. A typical example from the $\lambda$-calculus is
$\alpha$-equivalence, which identifies terms that differ only by the
names of bound variables. Strictly speaking, we represent a term by the set
of all terms that can be obtained from it by renaming variables. In order
to define a function on terms, we must show that the result is independent
of the choice of variable names. Here we see the drawbacks of
using equivalence classes: individuals are represented by sets and function
definitions carry a proof obligation.

Users of theorem proving tools are naturally inclined to prefer concrete
methods whenever possible. With their backgrounds in computer science, they
will see functions as algorithms and seek to find canonical representations
of elements. We can replace $\alpha$-equivalence by de Bruijn
variables~\cite{debruijn72}. Other common applications of equivalence
classes can similarly be replaced by
clever data structures. However, such methods can make proofs
unnecessarily difficult.

The automated reasoning community already uses equivalence classes. I
conducted an informal e-mail survey and learned of applications using
ACL2, Coq, HOL, Mizar and PVS\@. There must be other applications.
However, 
ACL2~\cite{acl2} does not use equivalence classes; it obtains a
similar effect through its ability to rewrite a term with respect to
any equivalence relation.%
\footnote{Documentation is on the Internet at
\url{http://www.cs.utexas.edu/users/moore/acl2/v2-7/EQUIVALENCE.html}}
The work in Coq is specific to constructive type theory and
uses the concept of setoid~\cite{geuvers-constructive}. The work using HOL requires the use of special purpose
code~\cite{harrison94,homeier-quotient}. If we consider the huge
amount of mathematics that has been formalised, it seems that uses of
equivalence classes are rare.

The object of this paper is to show that with the right definitions,
working with equivalence classes is easy. The approach will work in any
theorem prover that admits the formalisation of basic set theory. Function
definitions are made in a stylised but readable form. Once a function is
shown to respect the equivalence relation, further proofs can be undertaken
as they would be in a mathematics textbook: by assuming as given an
arbitrary representative of the equivalence class. Higher-order logic is
sufficient, by the obvious representation of typed sets by predicates;
untyped axiomatic set theory is also sufficient. The axiom of choice is
not required. The tool does not need to be programmable. The examples
in this paper have been done using Isabelle/HOL~\cite{isa-tutorial}.

The paper provides a brief review of equivalence classes
(\S\ref{sec:background}). It then outlines a formal lemma library for
equivalence classes (\S\ref{sec:Equiv}), which it demonstrates using
a construction of the integers (\S\ref{sec:IntDef}). It then
demonstrates how equational constraints can be imposed on a recursive
datatype (\S\ref{sec:quodatatype}). Finally, the paper
presents brief conclusions (\S\ref{sec:concl}).

\section{Equivalence Classes and Quotient Sets}\label{sec:background}

This section is a brief review of the mathematics required for this paper.
Thorough descriptions of the concepts appear in standard
textbooks~\cite[p.\ts128]{kolman-discrete}.

An \emph{equivalence relation} over a set is any relation that 
is reflexive, symmetric and transitive. Each of the following examples
is easily shown to be an equivalence relation.

\begin{itemize}
\item $\alpha$-equivalence has already been mentioned in the introduction.

\item The integers can be defined as equivalence classes
on pairs of natural numbers related by
$(x,y)\sim(u,v)$ if and only if $x+v=u+y$.

\item The rational numbers can be defined as equivalence classes on pairs
of integers related by $(x,y)\sim(u,v)$ if and only if $xv=uy$, where $y$,
$v\not=0$. Since $y$ and $v$ must be non-zero, it is an equivalence
relation on the set $\mathbb Z\times (\mathbb Z\setminus\{0\})$.
\end{itemize}

We can often do without equivalence classes.
The integers can be defined using a signed magnitude representation. The
rational numbers can be defined as fractions in reduced form. The drawbacks
of both representations become clear when we try to prove that addition is
associative. A signed magnitude representation requires case analysis on
the sign and the consideration of eight cases; proving the
associative law for reduced fractions will require a body of theory about
greatest common divisors~\cite[\S3]{harrison94}.

If ${\sim}$ is an equivalence relation over a set~$A$, and $x\in A$, then
$[x]_{\sim}$ denotes $\{y\mid y\sim x\}$ and is called an \emph{equivalence class}. The notation simplifies to $[x]$ when ${\sim}$ is fixed in advance.
Because ${\sim}$ is reflexive, an equivalence 
class is always non-empty. Because $\sim$ is symmetric and transitive, 
equivalence classes are disjoint: if two equivalence classes 
have an element in common, then they are equal. The equivalence 
relation therefore partitions the set~$A$. The quotient set $A/{\sim}$ is defined to be the set of equivalence classes generated by~${\sim}$.

Let us examine the construction of the integers. Here $A$ is the set of
pairs of natural numbers. As mentioned above, the equivalence relation
is $(x,y)\sim(u,v)$ if and only if $x+v=u+y$. The 
integer 0 and functions for negation, addition and multiplication are
defined in terms of equivalence classes. 
\begin{align*}
    0  &= [(0,0)] \\
    {}-[(x,y)] &= [(y,x)] \\
    [(x,y)] + [(u,v)] &= [(x+u,y+v)] \\
    [(x,y)] \times [(u,v)] &= [(xu+yv,xv+vu)]
\end{align*}
The intuition is that $[(x,y)]$ represents 
the integer $x-y$, so clearly the negation of  $[(x,y)]$ must be $[(y,x)]$.
The definition of multiplication is obtained by evaluating the product
$(x-y)(u-v)$. 

Such definitions are only legitimate if they are independent 
of the particular elements chosen from the equivalence classes. For example, negation is well-defined
because $[(x,y)]=[(x',y')]$ implies $-[(x,y)]=-[(x',y')]$. The statement that a function respects an equivalence relation is called a
\emph{congruence property}. For negation, it is easily verified: if
$(x,y)\sim(x',y')$, then $x+y'=x'+y$, so $y+x'=y'+x$ and thus
$(y,x)\sim(y',x')$. The congruence property for addition is harder  to
verify and that for multiplication is harder still. These proof obligations
are the chief drawback of using equivalence classes. Once we have verified
the congruence properties, developing the theory of the integers is easy.

To prove $-(-z)=z$, write the integer $z$ as $[(x,y)]$. (We have just
proved that the choice of $x$ and $y$ is irrelevant.) Now trivially
\[ -(-[(x,y)]) = -[(y,x)] = [(x,y)].\]
The proof that addition is associative appeals to the corresponding
property for the natural numbers:
\begin{align*}
 \bigl([(x_1,y_1)] + [(x_2,y_2)]\bigr) + [(x_3,y_3)] 
   &=[(x_1+x_2+x_3,y_1+y_2+y_3]) \\
   &= [(x_1,y_1)] + \bigl([(x_2,y_2)] + [(x_3,y_3)]\bigr).
\end{align*}

All of these proofs work in the same way. We write the given integers as
pairs of natural numbers. We simplify the expression using function
definitions that have already been shown to be legitimate. We are left with
elementary reasoning about the natural numbers. If we formalise equivalence
relations appropriately, the formal proofs will be as natural as those
shown above.

\section{A Formalization of Equivalence Classes}\label{sec:Equiv} 

The key to effective use of equivalence classes is to give definitions in a
particular form and to simplify them using particular lemmas. The
formalization is designed to make
the machine proofs as simple as possible, for any verifier. Unions capture
the possible ambiguity in a function defined on equivalence classes. For a
well-defined function, there will be no ambiguity.

This section
presents the most important components of the formalisation, omitting
obvious definitions and routine proofs. The approach should work with tools
such as HOL~\cite{mgordon-hol} and PVS~\cite{pvs-long-tutorial}. Isabelle
output is shown, using mathematical symbols that are largely
self-explanatory. Note that the Isabelle statement 
$\lbrakk P_1; \ldots; P_n\rbrakk \Rightarrow Q$
is equivalent to $P_1\Rightarrow \ldots\Rightarrow P_n \Rightarrow Q$
and denotes the inference rule $\frac{P_1\;\ldots\; P_n}{Q.}$

Another notation in need of explanation is the image operator. A relation in 
Isabelle/HOL is a set of ordered pairs. The notation $R``A$ denotes
the image of the set ~$A$ under the relation~$R$, namely $\{y\mid \exists x\in A\;(x,y)\in R\}$. In particular, $R``\{x\}$ denotes $\{y\mid (x,y)\in R\}$,
which is the equivalence class $[x]_R$ when $R$ is an equivalence
relation. (HOL users traditionally represent the relation~$R$ as a
function of type
$\alpha\to\alpha\to\isa{bool}$. They can express the equivalence class
$[x]_R$ by currying: $R\,x$. Either approach is acceptable.)
 
The predicate \isa{equiv\ A\ r} denotes that \isa{r} is an equivalence
relation on the set~\isa{A}\@. This concept is easily defined and its basic
properties proved. Let us therefore focus on the theorems that are directly
relevant to working with quotient types. This theorem, concerning equality
between equivalence classes, is useful in congruence proofs.
\begin{isabelle}
\isacommand{theorem}\ eq\_equiv\_class\_iff:\isanewline
\ \ "\isasymlbrakk equiv\ A\ r; \ x\ \isasymin \ A; \ y\ \isasymin \ A\isasymrbrakk\isanewline
\ \ \ \isasymLongrightarrow \ (r``\isacharbraceleft x\isacharbraceright \ =\ r``\isacharbraceleft y\isacharbraceright )\ =\
((x,y)\ \isasymin \ r)"
\end{isabelle}
The quotient set \isa{A//r} is defined to be the set of equivalence 
classes. 
\begin{isabelle}
\ \ "A//r\ \isasymequiv \ \isasymUnion x\ \isasymin \ A.\ \isacharbraceleft r``\isacharbraceleft x\isacharbraceright \isacharbraceright "\ \ \end{isabelle}
The combination of unions and singleton sets in this definition captures
the concept of a set comprehension. I have formalized
$\{f(x)\mid x\in A\}$ by $\bigcup_{x\in A}\{f(x)\}$. This technique
generalizes nicely: 
\[ \{f(x_1, \ldots, x_n)\mid x_1\in A_1, \ldots, x_n\in A_n\} =
   \bigcup_{x_1\in A_1}\ldots\bigcup_{x_n\in A_n}\{f(x_1, \ldots, x_n)\}. \]
I use the same technique below when defining functions over equivalence
classes. Nested unions of singleton sets work well. If an alternative
formalization is chosen, then the lemmas proved below about unions
must be reformulated accordingly.  
 
Crucial to the method is the
treatment of congruence properties.  The formula \isa{f\ respects\ r} expresses%
\footnote{Suggested by Rob Arthan. Previously, I used a less intuitive notation, \isa{\footnotesize congruent\ r\ f}.}
that the function \isa{f} respects the
relation~\isa{r}. That is, \isa{f} returns equal results for arguments
that are related by~\isa{r}.
\begin{isabelle}
f\ respects\ r\ \isasymequiv \ \isasymforall y\ z.\ (y,z)\ \isasymin \ r\ \isasymlongrightarrow \ f\ y\ =\ f\ z
\end{isabelle}
This definition does not constrain the domain and range of~\isa{f},
but in the lemmas involving unions, this function must range over
sets. In practice, these sets will either be singletons or
equivalence classes. In untyped set theory, where everything is a set,
they could be anything.

The next lemma expresses the central idea of the approach, although
users do not invoke it directly. It eliminates unions over constant
functions. For all values in its domain (namely~\isa{A}),
the function~\isa{f} returns the set~\isa{c}.
\begin{isabelle}
\isacommand{lemma}\ "\isasymlbrakk a\ \isasymin \ A; \ \isasymforall y\ \isasymin
\ A.\ f\ y\ =\ c\isasymrbrakk\ \isasymLongrightarrow \ (\isasymUnion y\ \isasymin \
A.\ f\ y)\ =\ c"
\end{isabelle}

The following lemma is again crucial. It eliminates a union over the
elements of an equivalence class provided the function respects the
equivalence relation. Such
unions will appear in the definitions of functions over quotient types. The
removal of the union is precisely the simplification we need to obtain
natural reasoning.
\begin{isabelle}
\isacommand{lemma}\ UN\_equiv\_class:\isanewline
\ \ \ \ \ "\isasymlbrakk equiv\ A\ r;\ f\ respects\ r;\ a\ \isasymin \ A\isasymrbrakk\isanewline
\ \ \ \ \ \ \isasymLongrightarrow \ (\isasymUnion x\ \isasymin \
r``\isacharbraceleft a\isacharbraceright .\ f\ x)\ =\ f\ a"
\end{isabelle}

For two-argument functions, we have the predicate \isa{congruent2}, where 
\isa{congruent2\ r1\ r2\ f} means the function 
\isa{f} respects the relations~\isa{r1} and~\isa{r2}. 
\begin{isabelle}
\ \ \ \isasymforall x\ y\ u\ v.\ (x,y)\ \isasymin \ r1\ \isasymlongrightarrow\ (u,v)\ \isasymin \ r2\ \isasymlongrightarrow \ f\ x\ u\ =\ f\ y\ v
\end{isabelle}
The intuitive syntax \isa{f\ respects2\ r} abbreviates the common situation 
when the two relations are identical, \isa{congruent2\ r\ r\ f}.

The congruence property for a two-argument function can be shown directly
from the definition. Occasionally, it is easier to show \isa{congruent2\
r1\ r2\ f} by showing that \isa{f} respects each relation separately.
It also suffices to show that \isa{f} is commutative and respects a
relation in one argument. These straightforward lemmas, based on
Harrison's HOL formalization, are omitted.

Here is a lemma to eliminate unions over the elements 
of equivalence classes, this time for two arguments. 
\begin{isabelle}
\isacommand{lemma}\ UN\_equiv\_class2:\isanewline
\ \ "\isasymlbrakk equiv\ A\ r1;\ equiv\ A\ r2;\ congruent2\ r1\ r2\ f;\ a1\ \isasymin \ A;\ a2\ \isasymin \ A\isasymrbrakk \isanewline
\ \ \ \isasymLongrightarrow \ (\isasymUnion x1\ \isasymin \ r1``\isacharbraceleft a1\isacharbraceright .\
\isasymUnion x2\ \isasymin \ r2``\isacharbraceleft a2\isacharbraceright .\ f\ x1\
x2)\ =\ f\ a1\ a2"
\end{isabelle}
It follows from the one argument case. The proof uses two lemmas 
concerning functions that satisfy \isa{congruent2\ r1\ r2\ f}. The first
lemma asserts that the one-argument function \isa{f\ a} 
obtained by currying, where \isa{a\ \isasymin \ A},  respects~\isa{r2}. 
The second lemma
asserts that \isa{\isasymUnion x2\ \isasymin \ r2``\isacharbraceleft
a2\isacharbraceright .\ f\ x1\ x2} respects~\isa{r1} when viewed as a
function of~\isa{x1}.

One further lemma is helpful. Isabelle~\cite[\S8.2.1]{isa-tutorial} and HOL
both allow a tuple of bound variables to appear wherever a bound variable
is expected. They are translated into a primitive un-curry operator. For
example, $\bigcup_{(x,y)\in A}\,B x y$ abbreviates $\bigcup_{z\in
A}\,(\lambda(x,y).\,B x y) z$, where $\lambda(x,y).\,B x y$ satisfies the
equation 
\[ (\lambda(x,y).\,B x y)(X,Y) = B X Y. \] 
The lemma rewrites the
left-hand side (which users may employ in definitions) into nested unions,
a form that allows the application of the union lemma. 
\begin{isabelle}
\isacommand{lemma}\ UN\_UN\_split\_split\_eq:\isanewline
\ \ \ "(\isasymUnion (x1,x2)\ \isasymin \ X.\ \isasymUnion (y1,y2)\ \isasymin \ Y.\ A\ x1\ x2\ y1\ y2)\ =\isanewline
\ \ \ \ (\isasymUnion x\isasymin X.\ \isasymUnion y\isasymin Y.\ (\isasymlambda (x1,x2).\ (\isasymlambda (y1,y2).\ A\ x1\ x2\ y1\ y2)\ y)\ x)"
\end{isabelle}
The lemma is unnecessary for people who are willing to use the projection
functions \isa{fst} and \isa{snd} (\isa{PROJ.1} and \isa{PROJ.2} in PVS)\@.
It is specific to the common case where the representing set~\isa{A} 
consists of ordered pairs.

\section{Example: Defining the Integers Formally} \label{sec:IntDef}
 
This section applies the lemma library to a formal development of the
integers. The technique is independent of particular theorem provers.
Isabelle's approach to type definition (which resembles HOL's)
clutters some of the formulae with abstraction and representation
functions. These will be absent in PVS or in any untyped formalism.

We also need a function \isa{contents} that returns the sole element of a
singleton set. It must satisfy the equation 
\isa{contents\ \isacharbraceleft x\isacharbraceright\ = x}. In higher-order logic, defining such a function
requires a definite description (the $\iota$-operator). It does not require
an indefinite description (Hilbert's $\epsilon$-operator) because we are
only interested in singleton sets.%
\footnote{The Isabelle tutorial
discusses description operators \cite[\S5.10]{isa-tutorial}.}

Standard set theories---untyped and without atoms---satisfy
$\bigcup(\{x\})=x$. We could therefore use $\bigcup$ for \isa{contents},
but we can do better still. Both \isa{contents} and singletons (the
inner
$\{\cdots\}$) can be eliminated from function definitions by virtue of
the equation 
\[ \bigcup\left(\bigcup_{x\in A}\,\{f(x)\}\right)=\bigcup_{x\in A}\,f(x). \]
Even in higher-order logic, this simplification can be used to define
some functions whose result involves a set construction. Indeed,
\isa{contents} and singletons can be removed from most of the
definitions below. This section uses the general form in order to
demonstrate the general technique; the next section will use
simplified definitions.

\subsection{Defining the Integers in Isabelle/HOL} 

We begin by defining the equivalence relation. 
\begin{isabelle}
\ \ \ \ "intrel\ \isasymequiv \ \isacharbraceleft ((x,y),(u,v))\ |\ x\ y\ u\ v.\ x+v\ =\ u+y\isacharbraceright "
\end{isabelle}
Next, we introduce the type int. 
\begin{isabelle}
\isacommand{typedef}\ (Integ)\ \ int\ =\ "UNIV//intrel"\isanewline
\ \ \isacommand{by}\ (auto\ simp\ add:\ quotient\_def)
\end{isabelle}
An Isabelle type definition equates a new type to a set.
The command \isa{\isacommand{by}\ (auto\ ...)} proves the set to be non-empty by appealing to the definition of quotients.
The set is given the name \isa{Integ} and is defined to be the set of pairs
of natural numbers quotiented by \isa{intrel}. The polymorphic constant
\isa{UNIV} denotes the universal set, here of type \isa{nat*nat}. The
representation function maps elements of type \isa{int} to elements of type
\isa{nat*nat} belonging to the set \isa{Integ}; the abstraction function
maps in the opposite direction.

Now we can define the integer constants 0 and~1.
\begin{isabelle}
\ \ "0\ \isasymequiv \ Abs\_Integ(intrel\ ``\ \isacharbraceleft (0,0)\isacharbraceright)"\isanewline
\ \ "1\ \isasymequiv \ Abs\_Integ(intrel\ ``\ \isacharbraceleft (1,0)\isacharbraceright)"
\end{isabelle}
Here \isa{intrel\ ``\ \isacharbraceleft (0,0)\isacharbraceright}
denotes the equivalence class $[(0,0)]$, which \isa{Abs\_Integ} coerces to
the new type~\isa{int}.

Now, we can do some preliminary proofs. A trivial equivalence is proved and
given to the simplifier (through the \isa{[simp]} annotation). It unfolds
membership assertions concerning the equivalence relation without unfolding
the definition in other contexts.
\begin{isabelle}
\isacommand{lemma}\ intrel\_iff[simp]:\ "(((x,y),(u,v))\ \isasymin \ intrel)\ =\ (x+v\ =\ u+y)"\isanewline
\isacommand{by}\ (simp\ add:\ intrel\_def)
\end{isabelle}

Proving that \isa{intrel} is an equivalence relation requires a one-line
simplifier call. (\isa{UNIV} again denotes the universal set of type
\isa{nat*nat}.)
\begin{isabelle}
\isacommand{lemma}\ equiv\_intrel:\ "equiv\ UNIV\ intrel"\isanewline
\isacommand{by}\ (simp\ add:\ intrel\_def\ equiv\_def\ refl\_def\ sym\_def\ trans\_def)
\end{isabelle}
A few other routine declarations (omitted) complete the setup of
type~\isa{int}. 

Reasoning about the integers requires a theorem 
stating that every integer is
represented by a pair of natural numbers. This theorem is trivially
expressed in first-order logic by $\forall z\;\exists x y\;z = [(x,y)]$. In
proofs, this theorem replaces given integer variables by equivalence
classes involving arbitrary natural numbers. Isabelle's natural deduction
framework can express this reasoning step directly as an inference rule.
\begin{isabelle}
\isacommand{lemma}\ eq\_Abs\_Integ\ [cases\ type:\ int]:\isanewline
\ \ \ \ \ "(\isasymAnd x\ y.\ z\ =\ Abs\_Integ(intrel``\isacharbraceleft (x,y)\isacharbraceright )\ \isasymLongrightarrow \ P)\ \isasymLongrightarrow \ P"
\end{isabelle}
The annotation \isa{cases type:\ int} informs Isabelle that case
analysis  on an integer variable---through the \isa{cases}
command---refers to this rule implicitly. This Isabelle-specific
feature should  not be difficult to imitate using other tools. In HOL,
it is easy to write a tactic that takes an integer variable, creates
a suitable instance of the theorem, and eliminates the  existential
quantifiers.

\subsection{A One-Argument Function on Equivalence Classes}

Unary minus illustrates the definition of functions on quotient types.
The idea is simple. We cannot pick an arbitrary element of an equivalence
class, but if the function respects the equivalence relation, 
then the choice of element does not
matter. Therefore, we form a set consisting of all values generated by all
elements of the equivalence class. This set will simplify to a singleton, 
whose value will be returned via the equation 
\isa{contents\ \isacharbraceleft x\isacharbraceright\ = x}.
\begin{isabelle}
\ "-z\ \isasymequiv \ contents\isanewline
\ \ \ \ \ \ \ \ (\isasymUnion (x,y)\isasymin Rep\_Integ z.\ \isacharbraceleft \ Abs\_Integ(intrel``\isacharbraceleft (y,x)\isacharbraceright )\ \isacharbraceright)"
\end{isabelle}
Here \isa{intrel\ ``\ \isacharbraceleft (y,x)\isacharbraceright} denotes
the equivalence class $[(y,x)]$. The argument of \isa{contents} is
the collection of all integers
$[(y,x)]$ such that $(x,y)$ belongs to the equivalence class for~$z$. This
collection will turn out to be a singleton.

\subsubsection{Proving the Characteristic Equation}

Let us apply these ideas to integer negation. We can prove its 
\emph{characteristic equation} ${}-[(x,y)] = [(y,x)]$, describing its behaviour on equivalence classes.
\begin{isabelle}
\isacommand{lemma}\ minus:\isanewline
\ \ "-\ Abs\_Integ(intrel``\isacharbraceleft (x,y)\isacharbraceright )\ =\ Abs\_Integ(intrel\ ``\ \isacharbraceleft (y,x)\isacharbraceright )"\isanewline
\isacommand{proof}\ -\isanewline
\ \ \isacommand{have}\ "(\isasymlambda (x,y).\ \isacharbraceleft Abs\_Integ\ (intrel``\isacharbraceleft (y,x)\isacharbraceright )\isacharbraceright )\ respects\ intrel"\isanewline
\ \ \ \ \isacommand{by}\ (simp\ add:\ congruent\_def)\ \isanewline
\ \ \isacommand{thus}\ ?thesis\isanewline
\ \ \ \ \isacommand{by}\ (simp\ add:\ minus\_int\_def\ UN\_equiv\_class\ [OF\ equiv\_intrel])\isanewline
\isacommand{qed}
\end{isabelle}
The first part of the proof concerns congruence: that of
the body of the union in the definition of negation. The
simplifier, given the definition of congruence (\isa{congruent\_def}),
immediately establishes that claim. In general, congruence properties
can be difficult to prove.

The second part of the proof
establishes the desired equation using the definition of negation
(\isa{minus\_int\_def}) and our theorem about unions over equivalence
classes. This part of the proof is always easy. Above,
\isa{UN\_equiv\_class\ [OF\ equiv\_intrel]}  denotes the instance of
\isa{UN\_equiv\_class} for the relation \isa{intrel}; since we have just
proved the necessary congruence property, it is in effect the following
rewrite rule:
\begin{isabelle}
\ \ \ \ \ "(\isasymUnion x\ \isasymin \
intrel``\isacharbraceleft a\isacharbraceright .\ f\ x)\ =\ f\ a"
\end{isabelle}
Using this rule, the left-hand side of the desired equation simplifies 
immediately to
\begin{isabelle}
\ "contents\ (\isacharbraceleft \ Abs\_Integ(intrel``\isacharbraceleft (y,x)\isacharbraceright )\ \isacharbraceright)"
\end{isabelle}
Then, because \isa{contents} is applied to a singleton, this simplifies in one step to the desired right-hand side. Even the most basic rewriting 
engine can perform the reasoning outlined above.

\subsubsection{An Alternative: the Axiom of Choice}

The obvious way to formalize equivalence classes, used by virtually all
other researchers, employs Hilbert's $\epsilon$-operator to choose a
representative of each equivalence class. In that approach, 
negation might be defined as follows:
\begin{isabelle}
\ "-z\ \isasymequiv \ (let\ (x,y) = choose (Rep\_Integ z)\ in\ Abs\_Integ(intrel``\isacharbraceleft (y,x)\isacharbraceright))"
\end{isabelle}
Here, \isa{choose} is a function that returns an arbitrary element of a
given set. This reliance on the axiom of choice does not lessen the proof
obligation: we must still show that the function respects the equivalence
relation. Reasoning about the axiom of choice will be difficult unless
we can find lemmas resembling \isa{UN\_equiv\_class} to eliminate the
operator \isa{choose}. Even if we can, the formalization involving unions
is preferable because it avoids a needless dependence on the axiom of 
choice.

\subsubsection{Reasoning about the Newly Defined Function}

Given the characteristic equation, proving properties of unary negation is
trivial. The approach is always the same. First, let the \isa{cases}
command replace integer variables by equivalence classes. Then,
call the simplifier to replace integer constants (such as 0) by their
definitions and to apply the characteristic equation and other
simplification rules. At this point, the formal proof steps duplicate those
of textbook proofs.

Consider the proof that negation is self-cancelling. 
\begin{isabelle}
\isacommand{lemma}\ zminus\_zminus:\ "-\ (-\ z)\ =\ (z::int)"\isanewline
\isacommand{apply}\ (cases\ z)\isanewline
\isacommand{apply}\ (simp\ add:\ minus)\isanewline
\isacommand{done}
\end{isabelle}
The \isa{cases} command leaves the following proof state: 
\begin{isabelle}
\ 1.\ \isasymAnd x\ y.\ z\ =\ Abs\_Integ\ (intrel\ ``\ \isacharbraceleft (x,\ y)\isacharbraceright )\ \isasymLongrightarrow \ -\ (-\ z)\ =\ z%
\end{isabelle}
The simplifier uses the characteristic equation (\isa{minus}) twice,
each time exchanging the two variables. The machine proof follows the 
informal one presented in \S\ref{sec:background} above.
\begin{isabelle}
\ \ -\ (-\ (Abs\_Integ\ (intrel\ ``\ \isacharbraceleft (x,\ y)\isacharbraceright)))\isanewline
\ \ \ \ =\ -\ (Abs\_Integ\ (intrel\ ``\ \isacharbraceleft (y,\ x)\isacharbraceright))\isanewline
\ \ \ \ =\ Abs\_Integ\ (intrel\ ``\ \isacharbraceleft (x,\ y)\isacharbraceright)
\end{isabelle}

\subsection{Two-Argument Functions on Equivalence Classes}

Addition and multiplication illustrate the 
treatment of two-argument functions. There are simply two unions 
instead of one. 
\begin{isabelle}
\ "z\ +\ w\ \isasymequiv \isanewline
\ \ \ \ \ contents\ (\isasymUnion (x,y)\isasymin Rep\_Integ z.\ \isasymUnion (u,v)\isasymin Rep\_Integ w.\isanewline
\ \ \ \ \ \ \ \ \ \ \ \ \ \ \ \isacharbraceleft \ Abs\_Integ(intrel``\isacharbraceleft (x+u,\ y+v)\isacharbraceright )\ \isacharbraceright )"\isanewline
\isanewline
\ "z\ *\ w\ \isasymequiv \isanewline
\ \ \ \ \ contents\ (\isasymUnion (x,y)\isasymin Rep\_Integ z.\ \isasymUnion (u,v)\isasymin Rep\_Integ w.\isanewline
\ \ \ \ \ \ \ \ \ \ \ \ \ \ \ \isacharbraceleft \ Abs\_Integ(intrel``\isacharbraceleft (x*u\ +\ y*v,\ x*v\ +\ y*u)\isacharbraceright )\ \isacharbraceright )"
\end{isabelle}

The characteristic equation for addition describes its effect 
on equivalence classes: $[(x,y)] + [(u,v)] = [(x+u,y+v)]$. 
The proof again begins by establishing congruence. 
Then the main theorem is established using the definition of addition
and our union theorems for two-argument functions. As with unary negation
above, the proof is a short and simple equational argument.
\begin{isabelle}
\isacommand{lemma}\ add:\isanewline
\ \ \ "Abs\_Integ\ (intrel``\isacharbraceleft (x,y)\isacharbraceright )\ +\ Abs\_Integ\ (intrel``\isacharbraceleft (u,v)\isacharbraceright )\ =\isanewline
\ \ \ \ Abs\_Integ\ (intrel``\isacharbraceleft (x+u,\ y+v)\isacharbraceright )"\isanewline
\isacommand{proof}\ -\isanewline
\ \ \isacommand{have}\ "(\isasymlambda z\ w.\ (\isasymlambda (x,y).\ (\isasymlambda (u,v).\isanewline
\ \ \ \ \ \ \ \ \ \isacharbraceleft Abs\_Integ\ (intrel\ ``\ \isacharbraceleft (x+u,\ y+v)\isacharbraceright )\isacharbraceright )\ w)\ z)\isanewline
\ \ \ \ \ \ \ \ respects2\ intrel"\isanewline
\ \ \ \ \isacommand{by}\ (simp\ add:\ congruent2\_def)\isanewline
\ \ \isacommand{thus}\ ?thesis\isanewline
\ \ \ \ \isacommand{by}\ (simp\ add:\ add\_int\_def\ UN\_UN\_split\_split\_eq\isanewline
\ \ \ \ \ \ \ \ \ \ \ \ \ \ \ \ \ \ UN\_equiv\_class2\ [OF\ equiv\_intrel])\isanewline
\isacommand{qed}
\end{isabelle}
The congruence property, \isa{...\ respects2\ intrel}, looks formidable.
However, manually applying the theorem \isa{UN\_equiv\_class2} displays the
required claim, which we can then paste into the proof script. The
arithmetic decision procedure for the natural numbers makes the proof
trivial.

Consider the proof that unary minus distributes over 
addition. 
\begin{isabelle}
\isacommand{lemma}\ zminus\_zadd\_distrib:\ "-\ (z\ +\ w)\ =\ (-\ z)\ +\ (-\ w::int)"\isanewline
\isacommand{apply}\ (cases\ z,\ cases\ w)\isanewline
\isacommand{apply}\ (simp\ add:\ minus\ add)\isanewline
\isacommand{done}\end{isabelle}

Here is the proof state after both applications of the \isa{cases} method. 
\begin{isabelle}
\ 1.\ \isasymAnd x\ y\ xa\ ya.\isanewline
\isaindent{\ 1.\ \ \ \ }\isasymlbrakk z\ =\ Abs\_Integ\ (intrel\ ``\ \isacharbraceleft (x,\ y)\isacharbraceright );\isanewline
\isaindent{\ 1.\ \ \ \ \ }w\ =\ Abs\_Integ\ (intrel\ ``\ \isacharbraceleft (xa,\ ya)\isacharbraceright )\isasymrbrakk \isanewline
\isaindent{\ 1.\ \ \ \ }\isasymLongrightarrow \ -\ (z\ +\ w)\ =\ -\ z\ +\ -\ w
\end{isabelle}
This subgoal is proved by rewriting using the characteristic equations 
for negation and addition. The formal reasoning is essentially 
the same as the textbook proof.

The treatment of multiplication is similar. The proof of congruence 
(omitted) requires some work because it lies outside the scope 
of linear arithmetic. Given that lemma, simple equational reasoning 
(as always) establishes the characteristic equation: 
\begin{isabelle}
\isacommand{lemma}\ mult:\isanewline
\ \ "Abs\_Integ((intrel``\isacharbraceleft (x,y)\isacharbraceright ))\ *\ Abs\_Integ((intrel``\isacharbraceleft (u,v)\isacharbraceright ))\ =\isanewline
\ \ \ Abs\_Integ(intrel\ ``\ \isacharbraceleft (x*u\ +\ y*v,\ x*v\ +\ y*u)\isacharbraceright )"\isanewline
\isacommand{by}\ (simp\ add:\ mult\_int\_def\ UN\_UN\_split\_split\_eq\ mult\_congruent2\isanewline
\ \ \ \ \ \ \ \ \ \ \ \ \ \ UN\_equiv\_class2\ [OF\ equiv\_intrel\ equiv\_intrel])
\end{isabelle}

We can now prove the standard theorems relating negation, addition and
multiplication. Each proof consists of \isa{cases} followed by
simplification with characteristic equations and the corresponding
properties of the natural numbers. Other proofs, not shown, are equally
trivial.
\begin{isabelle}
\isacommand{lemma}\ zmult\_zminus:\ "(-\ z)\ *\ w\ =\ -\ (z\ *\ (w::int))"\isanewline
\isacommand{by}\ (cases\ z,\ cases\ w,\ simp\ add:\ minus\ mult\ add\_ac)
\end{isabelle}

\begin{isabelle}
\isacommand{lemma}\ zmult\_assoc:\ "((z1::int)\ *\ z2)\ *\ z3\ =\ z1\ *\ (z2\ *\ z3)"\isanewline
\isacommand{by}\ (cases\ z1,\ cases\ z2,\ cases\ z3,\isanewline
\ \ \ \ simp\ add:\ mult\ add\_mult\_distrib2\ mult\_ac)
\end{isabelle}

\begin{isabelle}
\isacommand{lemma}\ zmult\_commute:\ "(z::int)\ *\ w\ =\ w\ *\ z"\isanewline
\isacommand{by}\ (cases\ z,\ cases\ w,\ simp\ add:\ mult\ add\_ac\ mult\_ac)\
\end{isabelle}

\begin{isabelle}
\isacommand{lemma}\ zadd\_zmult\_distrib:\ "((z1::int)+z2)\ *\ w\ =\ (z1*w)\ +\ (z2*w)"\isanewline
\isacommand{by}\ (cases\ z1,\ cases\ z2,\ cases\ w,\isanewline
\ \ \ \ simp\ add:\ add\ mult\ add\_mult\_distrib2\ mult\_ac)
\end{isabelle}

\subsection{Further Operations on the Integers}

The treatment of the ordering (${\le}$) illustrates 
an advantage of this approach. A relation in higher-order logic 
is a Boolean-valued function, so we could have used nested unions 
as we did for addition and multiplication. However, because we are using
native logic rather than a package, we are not 
forced to formalise this relation as a function. 
\begin{isabelle}
\ \ \ "z\ \isasymle \ (w::int)\ \isanewline
\ \ \ \ \isasymequiv \ \isasymexists x\ y\ u\ v.\ x+v\ \isasymle \ u+y\ \&\isanewline
\ \ \ \ \ \ \ \ \ \ \ \ \ \ \ \ \ (x,y)\ \isasymin\ Rep\_Integ\ z\ \&\ (u,v)\ \isasymin \ Rep\_Integ\ w"
\end{isabelle}

We can prove the characteristic equation directly, without proving 
congruence. The proof is trivial. Informally, the equation is $[(x,y)]\le[(u,v)]\iff x+v \le u+y$.
\begin{isabelle}
\isacommand{lemma}\ le:\isanewline
\ \ "(Abs\_Integ(intrel``\isacharbraceleft (x,y)\isacharbraceright )\ \isasymle \ Abs\_Integ(intrel``\isacharbraceleft (u,v)\isacharbraceright ))\isanewline
\ \ \ =\ (x+v\ \isasymle \ u+y)"\isanewline
\isacommand{by}\ (force\ simp\ add:\ le\_int\_def)
\end{isabelle}

The proofs about the ordering are largely straightforward, and 
are therefore omitted. The only difficult one is the monotonicity of multiplication, and it would be no easier using other treatments of
quotient types. 

A final example is the coercion from integers to natural numbers. 
It illustrates a function that leaves the integers.
\begin{isabelle}
\ \ \ \ "nat\ z\ \isasymequiv \ contents\ (\isasymUnion (x,y)\ \isasymin \ Rep\_Integ z.\ \isacharbraceleft \ x-y\ \isacharbraceright )"
\end{isabelle}
This function respects the equivalence relation, and its characteristic 
equation is $\mathrm{nat}[(x,y)] = x-y$.  Note that
$x-y$ is natural number subtraction and that $x-y=0$ if $x\le y$. 
\begin{isabelle}
\isacommand{lemma}\ nat:\ "nat\ (Abs\_Integ\ (intrel``\isacharbraceleft (x,y)\isacharbraceright ))\ =\ x-y"\isanewline
\isacommand{proof}\ -\isanewline
\ \ \isacommand{have}\ "(\isasymlambda (x,y).\ \isacharbraceleft x-y\isacharbraceright)\ respects\ intrel"\isanewline
\ \ \ \ \isacommand{by}\ (simp\ add:\ congruent\_def,\ arith)\ \isanewline
\ \ \isacommand{thus}\ ?thesis\isanewline
\ \ \ \ \isacommand{by}\ (simp\ add:\ nat\_def\ UN\_equiv\_class\ [OF\ equiv\_intrel])\isanewline
\isacommand{qed}
\end{isabelle}
Using this characteristic equation, theorems relating the function \isa{nat}
to the other integer operations are trivial to prove.

\section{Quotienting a Recursive Data type}\label{sec:quodatatype}
 
Another application of equivalence relations is to impose equations on recursive
datatypes. The necessary declarations are voluminous, but they are not
complicated and can be produced with the assistance of cut-and-paste. This section
presents an example inspired by cryptographic protocols. Many symmetric-key
cryptosystems provide separate decryption and encryption operations, which are
inverses of each other. Writing decryption of message~$X$ using key~$K$ as $D_K(X)$
and the corresponding encryption as $E_K(X)$, we have the equations
$D_K(E_K(X))=X$ and $E_K(D_K(X))=X$. Decryption can be applied to any
message, not just to the result of an encryption.

To define a datatype with equational constraints, first define an ordinary
datatype (which will be a free algebra). Then, define an equivalence
relation expressing the desired equations; the precise form is illustrated
below. Finally, quotient the datatype. The free datatype constructors are
easily lifted to the new recursive datatype, using the techniques of
function definition described above.  To define other functions on the new
datatype, first define a concrete version on the free datatype and then
lift it.

\subsection{The Concrete Datatype and the Equivalence Relation}

This simple datatype has four constructors: a message can be a nonce (a number),
the concatenation of two messages, an encryption or a decryption.
\begin{isabelle}
\isacommand{datatype}\isanewline
\ \ \ \ \ freemsg\ =\ NONCE\ \ nat\isanewline
\ \ \ \ \ \ \ \ \ \ \ \ \ |\ MPAIR\ \ freemsg\ freemsg\isanewline
\ \ \ \ \ \ \ \ \ \ \ \ \ |\ CRYPT\ \ nat\ freemsg\ \ \isanewline
\ \ \ \ \ \ \ \ \ \ \ \ \ |\ DECRYPT\ \ nat\ freemsg
\end{isabelle}

The equivalence relation, \isa{msgrel}, is defined inductively. The first
two rules (\isa{CD} and \isa{DC}) express the desired equations between
encryption and decryption. The next
four rules (\isa{NONCE} to \isa{DECRYPT}) have many purposes. They make the
equations hold for sub-messages; they allow the abstract constructors to
respect \isa{msgrel}; they ensure that \isa{msgrel} is reflexive. The last
two rules (\isa{SYM} and \isa{TRANS}) ensure that \isa{msgrel} is
symmetric and transitive.
\begin{isabelle}
\isacommand{inductive}\ "msgrel"\isanewline
\ \ \isakeyword{intros}\ \isanewline
\ \ \ \ CD:\ \ \ \ "CRYPT\ K\ (DECRYPT\ K\ X)\ \isasymsim \ X"\isanewline
\ \ \ \ DC:\ \ \ \ "DECRYPT\ K\ (CRYPT\ K\ X)\ \isasymsim \ X"\isanewline
\ \ \ \ NONCE:\ "NONCE\ N\ \isasymsim \ NONCE\ N"\isanewline
\ \ \ \ MPAIR:\ "\isasymlbrakk X\ \isasymsim \ X';\ Y\ \isasymsim \ Y'\isasymrbrakk \ \isasymLongrightarrow \ MPAIR\ X\ Y\ \isasymsim \ MPAIR\ X'\ Y'"\isanewline
\ \ \ \ CRYPT:\ "X\ \isasymsim \ X'\ \isasymLongrightarrow \ CRYPT\ K\ X\ \isasymsim \ CRYPT\ K\ X'"\isanewline
\ \ \ \ DECRYPT:\ "X\ \isasymsim \ X'\ \isasymLongrightarrow \ DECRYPT\ K\ X\ \isasymsim \ DECRYPT\ K\ X'"\isanewline
\ \ \ \ SYM:\ \ \ "X\ \isasymsim \ Y\ \isasymLongrightarrow \ Y\ \isasymsim \ X"\isanewline
\ \ \ \ TRANS:\ "\isasymlbrakk X\ \isasymsim \ Y;\ Y\ \isasymsim \ Z\isasymrbrakk \ \isasymLongrightarrow \ X\ \isasymsim \ Z"
\end{isabelle}
The relation~\isa{\isasymsim} is easily proved to be reflexive, symmetric 
and transitive. The proof of \isa{X\isasymsim X} is by structural induction on
the message~\isa{X}\@.

\subsection{Two Functions on the Free Algebra}\label{sec:free-functions}

Two examples will illustrate how functions are lifted to the quotiented abstract datatype. Obviously, we can only consider functions that
respect the equivalence relation. Both of these functions ignore encryption
and decryption altogether, so they are acceptable. 

The function \isa{freenonces} returns the set of all nonces present in a message.
It is defined by structural recursion.
\begin{isabelle}
\ \ \ "freenonces\ (NONCE\ N)\ =\ \isacharbraceleft N\isacharbraceright "\isanewline
\ \ \ "freenonces\ (MPAIR\ X\ Y)\ =\ freenonces\ X\ \isasymunion \ freenonces\ Y"\isanewline
\ \ \ "freenonces\ (CRYPT\ K\ X)\ =\ freenonces\ X"\isanewline
\ \ \ "freenonces\ (DECRYPT\ K\ X)\ =\ freenonces\ X"
\end{isabelle}
This function respects the equivalence relation.  The one-line proof 
appeals to induction on the definition of~\isa{\isasymsim} followed
by simplification.
\begin{isabelle}
\isacommand{theorem}\ msgrel\_imp\_eq\_freenonces:\ \isanewline
\ \ \ \ "U\ \isasymsim \ V\ \isasymLongrightarrow \ freenonces\ U\ =\ freenonces\ V"\isanewline
\isacommand{by}\ (erule\ msgrel.induct,\ auto)
\end{isabelle}

The function \isa{freeleft} returns the left part of the topmost \isa{MPAIR}
constructor.  (The lifted version, \isa{left}, will be a destructor function for the abstract
\isa{MPair} constructor.) The cases for \isa{CRYPT} and \isa{DECRYPT} make it
respect the equivalence relation. The case for \isa{NONCE} makes the 
function total, and will yield the further equation 
\isa{left\ (Nonce\ N)\ =\ Nonce\ N}.
\begin{isabelle}
\ \ \ "freeleft\ (NONCE\ N)\ =\ NONCE\ N"\isanewline
\ \ \ "freeleft\ (MPAIR\ X\ Y)\ =\ X"\isanewline
\ \ \ "freeleft\ (CRYPT\ K\ X)\ =\ freeleft\ X"\isanewline
\ \ \ "freeleft\ (DECRYPT\ K\ X)\ =\ freeleft\ X"
\end{isabelle}

The proof that \isa{freeleft} respects the equivalence relation resembles the
previous one, but includes an appeal to \isa{msgrel.intros}: a  list of
theorems for proving membership in~\isa{\isasymsim}.
\begin{isabelle}
\isacommand{theorem}\ "U\ \isasymsim \ V\ \isasymLongrightarrow \ freeleft\ U\ \isasymsim \ freeleft\ V"\isanewline
\isacommand{by}\ (erule\ msgrel.induct,\ auto\ intro:\ msgrel.intros)
\end{isabelle}

\subsection{The Abstract Message Type and its Constructors}

The abstract type of messages is declared by quotienting the universal
set (here of type \isa{freemsg}) with the relation~\isa{\isasymsim}.
\begin{isabelle}
\isacommand{typedef}\ (Msg)\ \ msg\ =\ "UNIV//msgrel"\isanewline
\ \ \isacommand{by}\ (auto\ simp\ add:\ quotient\_def)
\end{isabelle}

The abstract versions of the message constructors are called \isa{Nonce},
\isa{MPair}, \isa{Crypt} and \isa{Decrypt}. They are defined as functions
using unions, as we have already seen for the integer operations. The following
definitions do not use the function \isa{contents}; this simplification is possible
because the results are all derived from a set (namely an equivalence class).

\begin{isabelle}
\ "Nonce\ N\ ==\ Abs\_Msg(msgrel``\isacharbraceleft NONCE\ N\isacharbraceright )"\isanewline
\isanewline
\ "MPair\ X\ Y\ ==\isanewline
\ \ \ Abs\_Msg\ (\isasymUnion U\isasymin Rep\_Msg X.\ \isasymUnion V\isasymin Rep\_Msg Y.\ msgrel``\isacharbraceleft MPAIR\ U\ V\isacharbraceright )"\isanewline
\isanewline
\ "Crypt\ K\ X\ ==\ Abs\_Msg\ (\isasymUnion U\isasymin Rep\_Msg X.\ msgrel``\isacharbraceleft CRYPT\ K\ U\isacharbraceright )"\isanewline
\isanewline
\ "Decrypt\ K\ X\ ==\ Abs\_Msg\ (\isasymUnion U\isasymin Rep\_Msg X.\ msgrel``\isacharbraceleft DECRYPT\ K\ U\isacharbraceright )"
\end{isabelle}

Proving the characteristic equations for these constructors is
straightforward. Each equation relates an abstract constructor to the
corresponding concrete constructor. Each congruence proof is immediate by
the definition of the equivalence relation. Recall that the proof of the
characteristic equation from the congruence property is trivial.
\begin{isabelle}
\ "MPair\ (Abs\_Msg(msgrel``\isacharbraceleft U\isacharbraceright ))\ (Abs\_Msg(msgrel``\isacharbraceleft V\isacharbraceright ))\ =\ \isanewline
\ \ Abs\_Msg\ (msgrel``\isacharbraceleft MPAIR\ U\ V\isacharbraceright )"
\end{isabelle}

\begin{isabelle}
\ "Crypt\ K\ (Abs\_Msg(msgrel``\isacharbraceleft U\isacharbraceright ))\ =\ Abs\_Msg(msgrel``\isacharbraceleft CRYPT\ K\ U\isacharbraceright )"
\end{isabelle}

There is no characteristic equation for \isa{Nonce} because its argument is
not an equivalence class. The characteristic equation for \isa{Decrypt}
appears below, with its proof. As always, congruence is
established first.
\begin{isabelle}
\isacommand{lemma}\ Decrypt:\isanewline
\ \ \ \ \ "Decrypt\ K\ (Abs\_Msg(msgrel``\isacharbraceleft U\isacharbraceright ))\ =\isanewline
\ \ \ \ \ \ Abs\_Msg\ (msgrel``\isacharbraceleft DECRYPT\ K\ U\isacharbraceright )"\isanewline
\isacommand{proof}\ -\isanewline
\ \ \isacommand{have}\ "(\isasymlambda U.\ msgrel\ ``\ \isacharbraceleft DECRYPT\ K\ U\isacharbraceright)\ respects\ msgrel"\isanewline
\ \ \ \ \isacommand{by}\ (simp\ add:\ congruent\_def\ msgrel.DECRYPT)\isanewline
\ \ \isacommand{thus}\ ?thesis\isanewline
\ \ \ \ \isacommand{by}\ (simp\ add:\ Decrypt\_def\ UN\_equiv\_class\ [OF\ equiv\_msgrel])\isanewline
\isacommand{qed}
\end{isabelle}
As with the integers, the \isa{cases} lemma lets us replace an abstract
message by its representation as an equivalence class.
\begin{isabelle}
\isacommand{lemma}\ eq\_Abs\_Msg\ [cases\ type:\ msg]:\isanewline
\ \ \ \ \ "(!!U.\ z\ =\ Abs\_Msg(msgrel``\isacharbraceleft U\isacharbraceright )\ ==>\ P)\ ==>\ P"
\end{isabelle}
We now achieve a key objective: \isa{Crypt} and \isa{Decrypt}
are indeed inverses. Both proofs are one-liners using \isa{cases}, the
characteristic equations and the corresponding rule from the inductive
definition of~\isa{\isasymsim}.
\begin{isabelle}
\isacommand{theorem}\ CD\_eq:\ "Crypt\ K\ (Decrypt\ K\ X)\ =\ X"\isanewline
\isacommand{by}\ (cases\ X,\ simp\ add:\ Crypt\ Decrypt\ CD)\isanewline
\isanewline
\isacommand{theorem}\ DC\_eq:\ "Decrypt\ K\ (Crypt\ K\ X)\ =\ X"\isanewline
\isacommand{by}\ (cases\ X,\ simp\ add:\ Crypt\ Decrypt\ DC)
\end{isabelle}
Both proofs implicitly refer to the theorem \isa{eq\_equiv\_class\_iff},
presented in \S\ref{sec:Equiv} above, which relates equality of equivalence
classes to membership in the equivalence relation.

\subsection{Defining Functions on the Abstract Message Type}
 
To define a function on the abstract message type, first define an
analogous version on the concrete type and prove that it respects
equivalence relation. Then, define the abstract version as usual using
unions.
 
Recall that the function \isa{freenonces} returns the set of nonces contained in a
concrete message. We are now ready to declare the corresponding abstract function.
\begin{isabelle}
\ \ \ "nonces\ X\ ==\ \isasymUnion U\isasymin Rep\_Msg X.\ freenonces\ U"
\end{isabelle}
Congruence is immediate since \isa{freenonces} respects the equivalence relation,
as we saw in \S\ref{sec:free-functions} above.
\begin{isabelle}
\isacommand{lemma}\ nonces\_congruent:\ "freenonces\ respects\ msgrel"
\isanewline
\isacommand{by}\ (simp\ add:\ congruent\_def\ msgrel\_imp\_eq\_freenonces)
\end{isabelle}

The recursion equations for \isa{nonces}, the abstract function, are
trivial to prove. Here are the first three.
\begin{isabelle}
\ "nonces\ (Nonce\ N)\ =\ \isacharbraceleft N\isacharbraceright "\isanewline
\ "nonces\ (MPair\ X\ Y)\ =\ nonces\ X\ \isasymunion \ nonces\ Y"\isanewline
\ "nonces\ (Crypt\ K\ X)\ =\ nonces\ X"
\end{isabelle}
Here is the fourth equation, including its proof. Observe its similarity 
to proofs about the integer operations.
\begin{isabelle}
\isacommand{lemma}\ nonces\_Decrypt:\ "nonces\ (Decrypt\ K\ X)\ =\ nonces\ X"\isanewline
\isacommand{apply}\ (cases\ X)\ \isanewline
\isacommand{apply}\ (simp\ add:\ nonces\_def\ Decrypt\isanewline
\ \ \ \ \ \ \ \ \ \ \ \ \ \ \ \ \ UN\_equiv\_class\ [OF\ equiv\_msgrel\
nonces\_congruent])\isanewline
\isacommand{done}
\end{isabelle}

Recall that the function \isa{freeleft} returns returns the left part of the topmost
pair in a concrete message. The abstract version is defined using the techniques
just demonstrated.
\begin{isabelle}
\ \ \ "left\ X\ ==\ Abs\_Msg\ (\isasymUnion U\isasymin Rep\_Msg X.\ msgrel``\isacharbraceleft freeleft\ U\isacharbraceright )"
\end{isabelle}
Here are the recursion equations for this abstract destructor function.
\begin{isabelle}
\ "left\ (Nonce\ N)\ =\ Nonce\ N"\isanewline
\ "left\ (MPair\ X\ Y)\ =\ X"\isanewline
\ "left\ (Crypt\ K\ X)\ =\ left\ X"\isanewline
\ "left\ (Decrypt\ K\ X)\ =\ left\ X"
\end{isabelle}

\subsection{Freeness of the Abstract Constructors}
 
The abstract datatype satisfies equations between some of its constructors,
but the other constructors are still injective. In my experience, the
easiest way to prove such properties is to define explicit destructor
functions, not by inductive proofs over equivalence relations. We have seen
the function \isa{left} above, and the function \isa{right} can be defined
similarly. These two functions make it easy to prove that abstract message
pairing is injective.
\begin{isabelle}
\ "(MPair\ X\ Y\ =\ MPair\ X'\ Y')\ =\ (X=X'\ \&\ Y=Y')"
\end{isabelle}
The function \isa{nonces} makes it easy to prove that the abstract nonce constructor is injective. 
\begin{isabelle}
\ "(Nonce\ m\ =\ Nonce\ n)\ =\ (m\ =\ n)"
\end{isabelle}
Surprising as it may seem, the constructors for encryption and decryption
are also injective in the second argument; the proof is trivial, by the 
equations relating them. 
\begin{isabelle}
\ "(Crypt\ K\ X\ =\ Crypt\ K\ X')\ =\ (X=X')"\ \isanewline
\ "(Decrypt\ K\ X\ =\ Decrypt\ K\ X')\ =\ (X=X')"
\end{isabelle}
They are not injective in the first argument because any abstract message
can be rewritten to have the form of an encryption or decryption with
any desired key. For the same reason, we do not have the 
discrimination property \isa{Nonce\ N\ \isasymnoteq \ Crypt\ K\ X}.
However, many discrimination properties do hold, even for \isa{Crypt} and
\isa{Decrypt}. They can be proved by defining a discriminator function on 
the concrete datatype, using the methods demonstrated above. 
\begin{isabelle}
\ \ \ "freediscrim\ (NONCE\ N)\ =\ 0"\isanewline
\ \ \ "freediscrim\ (MPAIR\ X\ Y)\ =\ 1"\isanewline
\ \ \ "freediscrim\ (CRYPT\ K\ X)\ =\ freediscrim\ X\ +\ 2"\isanewline
\ \ \ "freediscrim\ (DECRYPT\ K\ X)\ =\ freediscrim\ X\ -\ 2"
\end{isabelle}
Because encryptions and decryptions cancel each other, this function 
respects the equivalence relation. Lifting this function to the quotiented
type yields a function \isa{discrim} satisfying the four analogous equations
on the abstract message constructors. Thus we can
prove that no nonce equals a message pair.
\begin{isabelle}
\ "Nonce\ N\ \isasymnoteq \ MPair\ X\ Y"
\end{isabelle}
We can also prove many discrimination results involving encryption, such as
this one.
\begin{isabelle}
\ "Crypt\ K\ (Nonce\ M)\ \isasymnoteq \ Nonce\ N"
\end{isabelle}

We have seen just one example of a quotiented datatype. Imposing
equations on the constructors seems to be straightforward, using the 
strategy shown above. Establishing other properties, such as inequations, 
requires defining suitable functions on the free datatype. 

At a referee's request, I have formalized an example involving nested
recursion: a quotiented datatype of expressions, including a constructor to
apply a function to a list of expressions. This development (available upon
request) uses a function to form a equivalence relation on lists from
an equivalence relation on list elements. It also requires a \isa{cases}
lemma analogous to \isa{eq\_Abs\_Msg}, but for lists of expressions. It is
clear that handling the full range of quotient datatypes requires
considerable ingenuity.

\section{Related work and Conclusions}\label{sec:concl}

The aim of this paper is to demystify the process of defining functions
over equivalence classes. Other authors have worked in this field.

Harrison~\citeyear[\S5]{harrison94} has written
an automated package for HOL that
declares the abstract quotient type and operations, returning theorems
about those operations. Its arguments include the equivalence relation, the
desired characteristic equations for each operation, and proofs of each
theorem expressed at the level of representatives. However, such a package
is not essential. The necessary definitions are straightforward, if they
are written as described above, and the reasoning about equivalence classes
poses no difficulties. Not using a package has its advantages: we
do not have to collect all the theorems we shall ever want into one giant
list; we are not restricted to top-level properties but can reason about
equivalence classes within a larger proof; we do not get stuck because the
package developer failed to anticipate our special requirements.

For defining abstract datatypes, the situation is different. Each
constructor and destructor function requires a separate declaration and
congruence proof. The declarations are uniform and the proofs are trivial,
but there are too many of them. Who would like to see the full treatment of
a 20 constructor abstract datatype? Nested and mutual recursion are also
difficult. If such declarations on needed
frequently, the proof tool should provide automated support.

Homeier~\citeyear{homeier-quotient} has developed an elaborate package for HOL
that offers special support for quotient constructions on recursive data
types. Homeier has used it to define a type of $\lambda$-terms quotiented
under $\alpha$-equivalence. Wenzel's treatment of quotient
types in Isabelle uses axiomatic type classes, which streamlines the notation but
tends to require additional type declarations.%
\footnote{See the theory at 
\url{http://isabelle.in.tum.de/library/HOL/Library/Quotient.html}.}
Slotosch~\citeyear{slotosch-quotients} has developed an
Isabelle/HOL theory of higher-order quotients based on  partial equivalence
relations.


PVS supports quotients in its standard prelude, but not with dedicated code.
Jackson has done a few examples, including a construction of the 
integers (private communication), while Tews~\citeyear{tews-lifting} has 
formalized one of the theorems in a theoretical paper on 
coalgebras. 

All of these other treatments of quotients use the axiom of choice,
typically via Hilbert's $\epsilon$-operator, to pick arbitrary
elements of equivalence classes. However, using the axiom of choice does
not lessen the proof obligations. Pragmatists  may argue that in
verification nobody cares whether choice is used or not. However,
pragmatists should be concerned that reasoning about $\epsilon$-terms is
tricky, while the unions over equivalence classes are simplified away
automatically.

We have seen, through a series of simple examples, how to define functions
on equivalence classes and how to reason about them. No special tools are
required, only a small lemma library. Each function definition must be
expressed in a particular form. Provided it respects the equivalence
relation, its characteristic equation is easily proved. Properties of the
function can then be reduced to properties of the representing type, with
proofs that resemble textbook presentations.

\begin{acks} Rob Arthan, Francis Flannery, Tom Harke,
Farhad Mehta, Lockwood Morris, Markus Wenzel 
and the referees commented on this paper.
Peter Homeier provided extensive information about his tools. The U.K.'s
Engineering and Physical Sciences Research Council (EPSRC) supported the
development of Isabelle.
\end{acks}

\bibliographystyle{acmtrans}\raggedright
\bibliography{string,atp,funprog,general,isabelle,theory,crossref}

\begin{received}
Received April 2004;
revised September 2004;
accepted September 2004
\end{received}
\end{document}